\documentclass[aps,prd,preprint,preprintnumbers,amsmath,amssymb,groupedaddress,nofootinbib]{revtex4}

\usepackage[]{graphicx}
\usepackage{amscd}
\usepackage[all,cmtip]{xy}
\usepackage{mathrsfs}
\usepackage{simplewick}
\usepackage{changepage}
\usepackage{bm}

\newcommand{\bea}{\begin{eqnarray}}
\newcommand{\eea}{\end{eqnarray}}
\newcommand{\beq}{\begin{equation}}
\newcommand{\eeq}{\end{equation}}

\newcommand{\De}{\Delta}
\newcommand{\la}{\lambda}
\newcommand{\de}{\delta}

\def\beq{\begin{equation}}
\def\eeq{\end{equation}}
\def\<{\langle}
\def\>{\rangle}

\def\cO {{\cal O}}

\def\Dp {{\Delta_{\phi}}}
\def\Ds {{\Delta_{\sigma}}}

\def\al{\alpha}

\begin{document}

%\preprint{MI-TH-1607}

\title{Bootstrapping Mixed Correlators in the Five Dimensional Critical $O(N)$ Models}

\author{ \vspace{8mm}Zhijin Li}
\email{lizhijin@physics.tamu.edu}
\author{Ning Su}
\email{suning1985@gmail.com}
\affiliation{{\small \vspace{4mm} George P. and Cynthia W. Mitchell Institute for
Fundamental Physics and Astronomy,
Texas A\&M University, College Station, TX 77843, USA} \vspace{16mm}}

\begin{abstract}
\vspace{3mm}
  We use the conformal bootstrap approach to explore $5D$ CFTs with $O(N)$ global symmetry, which contain $N$ scalars $\phi_i$ transforming as $O(N)$ vector. Specifically, we study multiple four-point correlators of the leading $O(N)$ vector $\phi_i$ and the $O(N)$ singlet $\sigma$. The crossing symmetry of the four-point functions and the unitarity condition provide nontrivial constraints on the scaling dimensions ($\Dp$, $\Ds$) of $\phi_i$ and $\sigma$. With reasonable assumptions on the gaps between scaling dimensions of $\phi_i$ ($\sigma$) and the next $O(N)$ vector (singlet) scalar, we are able to isolate the scaling dimensions $(\Dp$, $\Ds)$ in small islands. In particular, for large $N=500$, the isolated region is highly consistent with the result obtained from large $N$ expansion.
  We also study the interacting $O(N)$ CFTs for $1\leqslant N\leqslant100$. Isolated regions on  $(\De_\phi,\De_\sigma)$ plane are obtained using conformal bootstrap program with lower order of derivatives $\Lambda$; however, they disappear after increasing $\Lambda$. We think these islands are corresponding to interacting but nonunitary $O(N)$ CFTs. Our results provide a lower bound on the critical value $N_c>100$, below which the interacting $O(N)$ CFTs turn into nonunitary. The critical value is unexpectedly large comparing with previous estimations.
\end{abstract}

\maketitle

\newpage

\tableofcontents

\newpage

\section{Introduction}
The conformal bootstrap \cite{Ferrara:1973yt, Polyakov:1974gs, Mack:1975jr, Belavin:1984vu} provides a non-perturbative approach to solve conformal field theories (CFTs) using general consistency conditions of CFT. It has led to great successes in 2D, such as the seminal work \cite{Belavin:1984vu} on solving 2D rational CFTs. In recent years the conformal bootstrap has been revived since the breakthrough discovery in \cite{Rattazzi:2008pe}, which shows that the crossing symmetry and the unitary conditions can provide strong constraints on the operator scaling dimensions without an explicit form of Lagrangian. The crossing symmetry of four-point correlator leads to an infinite set of constraints on the CFT data. These constraints are difficult to be solved analytically, instead, they are truncated to a finite set and reformulated as a convex optimization problem so that they can be solved numerically. Here the convexity of conformal block functions \cite{Dolan:2000ut, Dolan:2003hv} plays a crucial role.
Since then the conformal bootstrap has been significantly developed and it becomes a remarkably powerful technique to obtain CFT data, including operator scaling dimensions and operator product expansion (OPE) coefficients in $D>2$ dimensions \cite{Rychkov:2009ij, Caracciolo:2009bx, Poland:2010wg, Rattazzi:2010gj, Rattazzi:2010yc, Vichi:2011ux, Poland:2011ey, ElShowk:2012ht, Liendo:2012hy, Beem:2013qxa, Kos:2013tga, El-Showk:2013nia, Alday:2013opa, Gaiotto:2013nva, Berkooz:2014yda, El-Showk:2014dwa, Nakayama:2014lva, Nakayama:2014yia, Alday:2014qfa, Chester:2014fya, Kos:2014bka, Caracciolo:2014cxa, Nakayama:2014sba, Golden:2014oqa, Chester:2014mea, Paulos:2014vya, Beem:2014zpa, Bae:2014hia, Chester:2014gqa, Simmons-Duffin:2015qma, Bobev:2015vsa, Kos:2015mba, Chester:2015qca, Beem:2015aoa, Iliesiu:2015qra, Poland:2015mta, Lemos:2015awa, Lin:2015wcg,
Chester:2015lej, Chester:2016wrc, Iha:2016ppj, Kos:2016ysd}. Review of previous developments on conformal bootstrap is provided in \cite{Simmons-Duffin:2016gjk}.

From conformal bootstrap with single correlator $\langle \phi\phi\phi\phi\rangle$, one can obtain bounds on the conformal
dimension or OPE coefficient of objective operator. The bounds may exhibit singular behaviors, such as kinks which are believed to be related to unitary CFTs.
One can expect to obtain more information on CFTs through
bootstrapping mixed correlators like $\langle\phi\phi\phi^2\phi^2\rangle$. Conformal bootstrap with mixed operators has been fulfilled in \cite{Kos:2014bka, Kos:2015mba} for $3D$ Ising model and critical $O(N)$ vector models
and the results are quite impressive---the allowed scaling dimensions are isolated in small islands. The accuracy can be improved further by refining the numerical techniques \cite{Simmons-Duffin:2015qma, Kos:2016ysd}. Studies on the $3D$ $O(N)$ vector models are strongly motivated by their special importance in physics. For small $N\leqslant3$ they describe second-order phase transitions happened in real physical systems \cite{Pelissetto:2000ek}. Besides, its $O(N)$-singlet sector is proposed to be dual to higher spin quantum gravity in $AdS_4$ with Dirichlet boundary conditions \cite{Klebanov:2002ja}. In the UV side, the $3D$ $O(N)$ vector model contains $N$ free scalars $\phi_i,~i=1, \cdots, N$ perturbed by quartic coupling $(\phi_i\phi_i)^2$. The RG flows to an IR fixed point which is strongly coupled. For the critical $O(N)$ vector models with large $N$ or in $D=4-\epsilon, \epsilon\ll1$ dimensions, one can obtain reliable results using large $N$ expansion or the well-known Wilson-Fisher $\epsilon$ expansion. Actually these analytical results have been used as consistency checks of conformal bootstrap in \cite{Kos:2013tga, El-Showk:2013nia}. Nevertheless, for the $3D$ ($\epsilon=1$) critical $O(N)$ vector models with small $N$ which are more physically attractive, these perturbative methods turn into less effective. In contrast, conformal bootstrap remains useful and has provided the most accurate results up to date \cite{Kos:2016ysd}.

Following the success of conformal bootstrap in critical $3D$ $O(N)$ vector models, one may expect to generalize the results to critical $O(N)$ vector models in higher dimensions. These models, if exist, are expected to provide examples on $\text{AdS}_{d+1}/\text{CFT}_d$ correspondence in higher dimensions. In 4D there is no critical $O(N)$ CFT, while in $D>4$, the interaction term $(\phi_i\phi_i)^2$ is irrelevant in the free $O(N)$ theory so the UV free $O(N)$ theory perturbed by the quartic interaction does not lead to an interacting fixed point in the IR, instead, the theory admits a Gaussian fixed point in the IR which flows to an interacting UV fixed point under $(\phi_i\phi_i)^2$ perturbation \cite{Parisi:1975im, Parisi:1977uz}. In $D=4+\epsilon$ such UV fixed point theory is weakly coupled for sufficient small $\epsilon$ and it requires a negative quartic coupling coefficient, which may introduce the problem of instability even though the scaling dimensions of the operators are above the unitary bound. A UV-completed formulation of the $O(N)$ model in $D>4$ dimensions has been proposed in \cite{Fei:2014yja, Fei:2014xta}
\begin{equation}
{\cal L}=\frac{1}{2}(\partial_\mu\phi_i)^2+\frac{1}{2}(\partial_i\sigma)^2+\frac{1}{2}g\sigma\phi_i^2+\frac{1}{6}\lambda\sigma^3, \label{lag}
\end{equation}
in which the $\phi_i$ constructs fundamental representation of $O(N)$ and the $O(N)$ singlet $\sigma$ performs as composite field $\phi_i^2$ in the UV side.
The theory contains cubic interaction terms which are relevant in space with dimension $D<6$.
Using the combination of $\epsilon$ and large $N$ expansion it has been shown that this theory admits an interacting IR fixed point \cite{Fei:2014yja, Fei:2014xta}, which is unitary for $N>N_c$, while below the critical value $N<N_c$ the coupling turns into complex and the IR fixed point theory is nonunitary. At one-loop level the critical value $N_c$ is about $N_c\approx1038$.
For $5D$ ($\epsilon=1$) critical $O(N)$ theories, the small $\epsilon$ condition for $\epsilon$ expansion approach breaks down so the results obtained from $\epsilon$ expansion should be treated carefully. Actually the critical value decreases to $N_c\approx64$ at three-loop level. In \cite{Gracey:2015tta} the author has obtained a critical value $N_c\approx400$ at four-loop level based on resummation methods. A non-perturbative method is desirable to determine the critical value $N_c$ in $5D$. The $5D$ critical $O(N)$ models have been studied using the nonperturbative functional renormalization group equations \cite{Rosten:2008ts, Percacci:2014tfa, Mati:2014xma, Kamikado:2016dvw, Eichhorn:2016hdi}, and Refs \cite{Rosten:2008ts, Percacci:2014tfa, Mati:2014xma, Kamikado:2016dvw} that no interacting fixed point exists at large N, while the analysis in \cite{Eichhorn:2016hdi} agrees with the results from the $D=6-\epsilon$ perturbative
approach when $\epsilon\ll1$ and predicts the $5D$ critical value $N_c=1$. The stability problem of IR fixed point in the cubic model remains according to \cite{Eichhorn:2016hdi}.

The conformal bootstrap approach has been employed to study $5D$ critical $O(N)$ models in \cite{Nakayama:2014yia, Bae:2014hia, Chester:2014gqa} following the proposal of the cubic model \cite{Fei:2014yja, Fei:2014xta}. In \cite{Nakayama:2014yia} the $5D$ critical $O(N)$ models have been assumed to saturate the minimum of the $O(N)$ current central charge $c_J$ for large $N$ and the existence of $5D$ critical $O(N)$ models are indicated from these minimums obtained from conformal bootstrap. The authors focused on bootstrapping the OPE coefficients rather than the scaling dimensions of conformal primary operators.
In $3D$ conformal bootstrap the interacting $O(N)$ CFTs have been found to lie at the kinks of the bounds for the scaling dimension $\Delta_\sigma$ of the $O(N)$ singlet $\sigma$, which appears as lowest dimension operator in the $S$ channel of the correlator $\langle\phi_i\phi_j\phi_k\phi_l\rangle$. However, in $5D$ cubic model the lowest dimension $O(N)$ singlet operator $\sigma$ performs as $\phi_i^2$, $\Delta_\sigma=2\Delta_{\phi}=3$  at the UV Gaussian fixed point which reduces to $\Delta_\sigma=2+O(1/N)$ near the IR fixed point. The IR fixed point is below the upper bound of scaling dimensions $\Delta_\sigma$ so there is no clue on the fixed point theory in the bound of scaling dimensions. This problem has been overcome in \cite{Bae:2014hia, Chester:2014gqa} by imposing a
gap on the scaling dimensions of $\sigma$  and the second lowest $O(N)$ singlet conformal primary scalar. With a reasonable assumption on the gap, the allowed region of the scaling dimensions $(\Delta_{\phi}, \Delta_\sigma)$ can be carved out and forms two sharp kinks. The UV Gaussian fixed point lies at the higher kink while the lower kink agrees with the large $N$ expansion predictions on IR interacting fixed point theories. Furthermore, the kink disappears for small $N\approx15$ which may indicate a small critical value $N_c$ \cite{Chester:2014gqa}.

However, one should be careful to consider the kinks in conformal dimension bound or the minimum of central charges as unitary CFTs. From perturbative methods it is known that in $D=6-\epsilon,~\epsilon\ll1$ the IR fixed point of cubic $O(N)$ models is endowed with complex critical couplings for $N\leqslant1000$. \footnote{CFTs in fractional dimensions are known to be nonunitary even with real couplings \cite{Hogervorst:2014rta, Hogervorst:2015akt}. However, the unitarity is violated by operators with high scaling dimensions so they are more difficult to be tested through conformal bootstrap approach.} Nevertheless, in \cite{Chester:2014gqa} a sharp kink is still generated from conformal bootstrap for $D=5.95$, $N=600$ which is much lower than the threshold value and should be nonunitary. The reason seems to be that the precision adopted in \cite{Chester:2014gqa} is not high enough to detect the small violation of unitary. A more powerful bootstrap approach is needed to study the $5D$ $O(N)$ models, especially on its critical value $N_c$.

In this work, we will study the conformal bootstrap with multiple correlators of conformal primaries $\phi_i$ and $\sigma$: $\langle\phi_i\phi_j\phi_k\phi_l\rangle$, $\langle\phi_i\phi_j\sigma\sigma\rangle$, $\langle\sigma\sigma\sigma\sigma\rangle$. Since there are more operators involved in the bootstrap program, it is
expected that the results will provide more rigid restrictions on the scaling dimensions of $(\Delta_{\phi}, \Delta_\sigma)$. Actually we find that the scaling dimensions $(\Delta_{\phi}, \Delta_\sigma)$ obtained from bootstrapping multiple correlators of $5D$ $O(500)$ model is isolated in a rather small island, which is nicely compatible with the perturbative results. We also study the critical value $N_c$ in $5D$. In preliminary numerical calculations we find small islands on the allowed scaling dimensions $(\Delta_{\phi}, \Delta_\sigma)$ for all $N\geqslant1$. However, these islands disappear after improving the bootstrapping precisions. Taking $N\approx40$ for example, it shows an apparent kink in the bound from bootstrapping single correlator of $\phi_i$s \cite{Chester:2014gqa}. Using multiple correlator conformal bootstrap with small $\Lambda$, we obtain an island on $(\Delta_{\phi}, \Delta_\sigma)$ plane close to the kink presented in \cite{Chester:2014gqa}, while it vanishes after increasing $\Lambda$. We do not find a stable island even for $N=100$, therefore our results suggest a rather large critical value $N_c>100$.

This paper is organized as follows. In section 2 we briefly review the cubic model of $O(N)$ vector model in $4<D<6$ and the perturbative results on scaling dimensions of lowest primary scalars. The scaling dimensions $(\Delta_{\phi}, \Delta_\sigma)$ obtained from large $N$ and $\epsilon$ expansions provide consistency checks for the results from conformal bootstrap. In section 3 we introduce the numerical conformal bootstrap equations for $5D$ $O(N)$ vector models and their numerical implementation. Our results are presented in section 4. We show that through bootstrapping the multiple correlators the scaling dimensions $(\Delta_{\phi}, \Delta_\sigma)$ are isolated in a small island for large $N=500$, while disappear with large $\Lambda$ for $N\leqslant100$. Conclusions are made in section 5.

\section{Perturbative Results for $5D$ Critical $O(N)$ Models}
The critical $O(N)$ vector model with quartic interaction in arbitrary dimensions $D=4-\epsilon$
 has been analyzed using the large $N$ expansions \cite{Vasiliev:1981yc, Vasiliev:1981dg, Vasiliev:1982dc, Lang:1990ni, Lang:1991kp, Lang:1992zw, Lang:1992pp,
Petkou:1994ad, Petkou:1995vu, Broadhurst:1996ur, Gracey:2002qa}. In $2<D<4$ ($\epsilon>0$), the quartic interaction is relevant and the RG flows from UV Gaussian fixed point to interacting IR fixed point perturbed by this coupling. The quartic interaction is irrelevant in $4<D<6$ ($\epsilon<0$) so the long-range physics is described by free field theory. The quartic coupling generates RG flow from the IR Gaussian fixed point to an interacting UV fixed point. The perturbative result for small $\epsilon$ shows the interaction coupling is negative at interacting UV fixed point which may lead to the stability problem. However, the scaling dimensions of scalar operators obtained from the large $N$ expansion are still above unitary bound and the unitary conditions remain unbroken for sufficient large $N$. One may expect the interacting UV fixed point from quartic model describes a universality class with $O(N)$ global symmetry in $4<D<6$
whose stable or metastable formulation may be realized in different model.

In $D=5$ spacetime, the conformal dimensions of $\phi_i$ and $\sigma$ have been evaluated at three-loop level
\bea
\Delta_{\phi}&=&\frac{3}{2}+\frac{0.216152}{N}-\frac{4.342}{N^2}-\frac{121.673}{N^3}+\cdots \label{Dphi}\\
\Delta_\sigma&=&2+\frac{10.3753}{N}+\frac{206.542}{N^2}+\cdots \label{Dsigma}\\
\Delta_{\sigma^2}&=& 4-\frac{13.8337}{N}-\frac{1819.66}{N^2}+\cdots \label{Dsigma2}
\eea
According to above $1/N$ expansion, the conformal dimension of $\phi_i$ is above the unitary bound ($\Delta_{\phi}>3/2$ for scalar fields) given $N>35$. The critical value $N_c=35$ can be significantly modified by by higher order corrections. Actually
the $5D$ $1/N$ expansions converge much slower than those in $3D$ \cite{Fei:2014yja}.

Alternatively, the $5D$ quartic theory can also be studied using $\epsilon$ expansion \cite{Wilson:1973jj}.  Conformal dimensions of $\phi_i$ and $\phi^2$ ($\sigma$)
have been calculated up to five-loop  \cite{Kleinert:1991rg}:
\beq
\Delta_{\phi}=1-\frac{\epsilon}{2}+\frac{N+2}{4(N+8)^2}\epsilon^2(1+a_1\epsilon+a_2\epsilon^2+a_3\epsilon^3), \label{Dphie}
\eeq
where
{\small
\bea
a_1&=& \frac{-N^2+56N+272}{4(N+8)^2}, \nonumber \\
a_2&=& -\frac{1}{16(N+8)^4}(5N^4+230N^3-1124N^2-17920N-46144+384\zeta(3)(N+8)(5N+22)), \nonumber \\
a_3&=& -\frac{1}{64(N+8)^6}\left(13N^6+946N^5+27620N^4+121472N^3-262528N^2-2912768N-5655552 \right.\nonumber \\
&&  \left. -16\zeta(3)(N+8)(N^5+10N^4+1220N^3-1136N^2-68672N-171264) \right.\nonumber \\
&&  \left. +1152\zeta(4)(N+8)^3(5N+22)-5120\zeta(5)(N+8)^2(2N^2+55N+186)\right), \nonumber
\eea}
and
\bea
\De_\sigma=2-\epsilon+\frac{N+2}{N+8}\epsilon\left(1+c_1\epsilon+c_2\epsilon^2+c_3\epsilon^3+c_4\epsilon^4\right), \label{Dsigmae}
\eea
where {\small
\bea
c_1&=& \frac{13N+44}{2(N+8)^2}, \nonumber \\
c_2&=& -\frac{1}{8(N+8)^4} (3 N^3-452 N^2-2672 N-5312 +96\zeta(3) (N+8) (5 N+22) ),\nonumber\\
c_3&=& -\frac{1}{32(N+8)^6}\left(3 N^5+398 N^4-12900 N^3 -81552 N^2 -219968 N -357120 \right.\nonumber \\
&& \left. +16\zeta(3)(N+8)(3 N^4-194 N^3+148 N^2+9472 N+19488) \right. \nonumber \\
&& \left.+288\zeta (4) (N+8)^3 (5 N+22) -1280\zeta (5) (N+8)^2 (2 N^2+55 N+186)\right), \nonumber \\
%%%%%%%%%%%%%%%
c_4&=&-\frac{1}{128(N+8)^8}\times \nonumber \\
&&\left(3 N^7-1198 N^6-27484 N^5-1055344 N^4-5242112 N^3-5256704 N^2+6999040 N-626688 \right.\nonumber \\
&& \left.-16\zeta(3)(N+8)(19004 N^4+102400 N^3+13 N^6-310 N^5-381536 N^2-2792576 N-4240640) \right. \nonumber \\
&& \left. -1024\zeta (3)^2 (N+8)^2 (2 N^4+18 N^3+981 N^2+6994 N+11688) \right.\nonumber \\
&& \left.+48 \zeta (4) (N+8)^3 (148 N^2+ 3 N^4-194 N^3 +9472 N+19488) \right.\nonumber \\
&& \left.+256\zeta(5) (N+8)^2 (155 N^4+3026 N^3+989 N^2-66018 N-130608) \right. \nonumber  \\
&& \left. -6400\zeta(6)(2 N^2+55 N+186) (N+8)^4 +56448\zeta (7) (14 N^2+189 N+526) (N+8)^3 \right). \nonumber
\eea}
Taking $\epsilon=-1$ the results can be interpolated to $5D$. For large $N$ the higher order coefficients
$c_i$s are of order $1/N$. In this case the $\epsilon$ expansion performs worse asymptotically in $5D$ comparing with the large $N$ expansion. While for small $N$
it is not clear at this stage which approach can provide better estimation. These perturbative results will be useful to estimate the conformal dimension gap which can be applied in the conformal bootstrap to improve the
numerical efficiency.

Both the large $N$ expansion and the $\epsilon$ expansion contain negative terms at higher loop level. For small $N$s these negative contributions may play dominating roles in the perturbative expansion and result in negative anomalous dimension. Specifically the five-loop result (\ref{Dphie}) shows the conformal dimension $\Delta_{\phi}<3/2$ for $N\leqslant14$ \cite{Fei:2014yja}.
In \cite{Chester:2014gqa} the conformal bootstrap with single correlator has been applied to generate bound on $\De_\sigma$. Interestingly the bounds are featured with kinks which are expected to relate to certain unitary fixed point theories while the kinks disappear near $N\approx15$, close to
the critical value estimated from $\epsilon$ expansion. However, as in the large $N$ expansion, the $\epsilon$ expansion in $5D$ is not converged up to fifth order and the contributions from higher loops are likely to modify the threshold value $N_c$ significantly.

The cubic $O(N)$ model (\ref{lag}) provides an approach to realize stable interacting $O(N)$ fixed point in $5D$ \cite{Fei:2014yja, Fei:2014xta}. The authors show that at one of the IR fixed point the cubic $O(N)$ model shares the same relevant critical exponents with the quartic $O(N)$ model so the two models are expected to describe the same universality class.\footnote{The renormalization group approach suggests the cubic model admits an extra RG relevant direction with positive critical exponent at the IR fixed point \cite{Eichhorn:2016hdi}. In this sense the universality class of the quartic $O(N)$ model is a subset of that of cubic $O(N)$ model.}
Like the quartic $O(N)$ model, the cubic $O(N)$ model also requires a critical value $N_c$ from unitarity constraint. In the cubic model, the unitarity is violated
in the way that the coupling coefficients acquire imaginary part when $N<N_c$. In \cite{Fei:2014yja, Fei:2014xta} the critical value $N_c$ is evaluated up to order $\epsilon^2$ in arbitrary dimension $D=6-\epsilon$. Four-loop results which include corrections on $N_c$ at order $\epsilon^3$ have been calculated in \cite{Gracey:2015tta}
\beq
N_c=1038.26605-609.83980\epsilon-364.17333\epsilon^2+452.71060\epsilon^3+O(\epsilon^4).
\eeq
As usual, above perturbative result is not sufficient to make a solid estimation on $5D ~(\epsilon=1)$ $N_c$ due to its asymptotic performance. It is tempting to evaluate the critical value $N_c$ using non-perturbative method.
 Besides the above interacting IR fixed point, the cubic model also admits extra fixed points with different critical value $N_c^\prime$; however, they are not corresponding to the classical interacting quartic fixed point and will not be studied in this work.

\section{Conformal Bootstrap  with Multiple Correlators}
Conformal bootstrap with multiple correlators has been developed in \cite{Kos:2014bka, Kos:2015mba} which aimed to solve the $3D$ Ising model and $O(N)$ vector model. This approach  has obtained the most accurate solutions on $3D$ Ising model and $O(N)$ vector model up to date \cite{Kos:2016ysd}.
Here we briefly introduce the conformal bootstrap program for $5D$ $O(N)$ vector model analogous to that for $3D$ $O(N)$ vector model \cite{Kos:2015mba}. More details on this program are provided in \cite{Simmons-Duffin:2015qma}.

\subsection{Bootstrap Equations from Crossing Symmetry}
Conformal partial wave function
 is the crucial ingredient for conformal bootstrap. In even dimensions $D=2,4,6$, the conformal partial wave functions have
been solved analytically \cite{Dolan:2000ut, Dolan:2003hv}. In odd dimensions there is no analytical expression for conformal partial wave function; however,
it can be calculated recursively with arbitrary precision \cite{Kos:2013tga, Kos:2014bka, Penedones:2015aga}. \footnote{Details on calculating conformal block function in arbitrary dimensions are provided in \cite{Paulos:2014vya} as part of an open-source numerical conformal bootstrap program JuliBootS. In this work we will use the JuliBoots code to calculate the conformal block functions of scalar operators in $5D$.}
 The general four-point function of scalar
operators can be expanded
in terms of conformal partial waves
\beq
\langle \sigma_1\sigma_2\sigma_3\sigma_4\rangle=\frac{1}{x_{12}^{\De_1+\De_2}x_{34}^{\De_3+\De_4}}\left(\frac{x_{24}}{x_{14}}\right)^{\De_{12}}
\left(\frac{x_{14}}{x_{13}}\right)^{\De_{34}}\sum_{\cO}\la_{12\cO}\la_{34\cO}g_{\De,\ell}^{\De_{12},\De_{34}}(u,v),
\eeq
where $\sigma_i$s are scalar operators with conformal dimension $\De_i$ ($\De_{ij}=\De_i-\De_j$) and $\cO$ is the conformal primary operator appears in the
OPE expansion of $\sigma_1\sigma_2\sim\la_{12\cO}\cO$ (and also $\sigma_3\sigma_4\sim\la_{34\cO}\cO$), whose conformal dimension and spin are $(\De, ~\ell)$.
The conformal invariant cross ratios $u,~ v$ are of the standard form $u=\frac{x_{12}^2x_{34}^2}{x_{13}^2x_{24}^2}$ and $v=\frac{x_{14}^2x_{23}^2}{x_{13}^2x_{24}^2}$,
$x_{ij}=|x_i-x_j|$.

The four-point function can be evaluated equivalently in different channels, as suggested by crossing symmetry, and it leads to the following equations
\beq
\sum_\cO\left(\la_{12\cO}\la_{34\cO}F_{\mp,\De,\ell}^{12,34}(u,v)\pm\la_{32\cO}\la_{14\cO}F_{\mp,\De,\ell}^{32,14}(u,v) \right)=0, \label{cross}
\eeq
in which
\beq
F_{\mp,\De,\ell}^{12,34}(u,v)=v^{\frac{\De_2+\De_3}{2}}g_{\De,\ell}^{\De_{12},\De_{34}}(u,v)\mp u^{\frac{\De_2+\De_3}{2}}g_{\De,\ell}^{\De_{12},\De_{34}}(v,u).
\eeq
To study the $5D$ $O(N)$ vector model, we apply the crossing relations for correlators $\langle\phi_i\phi_j\phi_k\phi_l\rangle$, $\langle\sigma\sigma
\sigma\sigma\rangle$ and $\langle\phi_i\phi_j\sigma\sigma\rangle$.
The $O(N)$ indices in the correlators are decomposed into three irreducible structures:
the $O(N)$ invariant, traceless symmetric and antisymmetric tensors. The conformal primaries appearing
in the OPE of $O(N)$ vector representations $\phi_i$
can be classified into three irreducible representations:
 \beq
 \phi_i\times\phi_j\sim \sum_{S}\la_{\phi\phi\cO_S}\cO\de_{ij}+\sum_{T}\la_{\phi\phi\cO_T}\cO_{(ij)}+\sum_{A}\la_{\phi\phi\cO_A}\cO_{[ij]},
 \eeq
 in which $S,~ T$ and $A$ denote $O(N)$ singlet, traceless symmetric tensor and anti-symmetric tensor representations.
 Consequently, the four-point correlator $\langle\phi_i\phi_j\phi_k\phi_l\rangle$ and its crossing symmetric partners are separated into three channels: $S, T, A$.
 For the mixed four-point correlator $\langle\phi_i\sigma\phi_j\sigma\rangle$, one needs to consider the OPE $\phi_i\sigma\sim\sum_{V}\la_{\phi_i\sigma\cO_i}\cO_i$
 which introduces the vector representations (denoted by $V$) as propagating operators in the mixed four-point correlator and its crossing symmetric partner.

The crossing relations for bootstrapping $5D$ $O(N)$ critical theories are essentially the same as those for $3D ~O(N)$ vector model \cite{Kos:2015mba}.
These equations can be written in a compact form \cite{Kos:2015mba} which are presented below for later reference
\bea
0 &=& \sum_{\cO_S} \left(\begin{array}{ccc} \la_{\phi\phi \cO_S} & \la_{\sigma\sigma \cO_S} \end{array} \right) \vec{V}_{S,\De,\ell}
\left( \begin{array}{c} \la_{\phi\phi \cO_S} \\ \la_{\sigma\sigma \cO_S} \end{array} \right) \nonumber\\
&&+\sum_{\cO_T} \la_{\phi\phi \cO_T}^2
\vec{V}_{T,\De,\ell}+ \sum_{\cO_A} \la_{\phi\phi \cO_A}^2 \vec{V}_{A,\De,\ell}+ \sum_{\cO_V} \la_{\phi \sigma \cO_V}^2 \vec{V}_{V,\De,\ell}. \label{booteq}
\eea
Explicit forms of the 7-component vectors $\vec{V}_S, \vec{V}_{T}, \vec{V}_A, \vec{V}_V$ are provided in the Appendix.

\subsection{Bounds from Crossing Relations}
The equations from crossing symmetry (\ref{booteq}) provide nontrivial constraints on the CFT data.
The numerical approach to study these equations was first proposed in \cite{Rattazzi:2008pe} and the following developments show this method is extremely powerful.
The logic of numerical conformal bootstrap is firstly to make assumptions on the CFT spectra.
If the assumptions are physical they are required to satisfy the crossing relations (\ref{booteq}) and the unitary condition. Numerical conformal bootstrap provides a systematical way to check the consistency between the assumptions and general constraints on CFTs. Bounds on the CFT data, including conformal dimensions of primary operators and OPE coefficients can be obtained by falsifying possible assumptions on the CFT spectra.

Specifically for any hypothetical spectra $(\De_\phi,\De_\sigma)$ above the unitary bounds,
they should be consistent with the crossing relations
(\ref{booteq}). However, if there are linear functionals $\vec{\al}=(\al_1,\al_2,\cdots,\al_7)$ satisfying
\bea
(\begin{array}{cc}1 & 1\end{array})~ \vec{\al}&\cdot & \vec{V}_{S,0,0}\left(\begin{array}{c}1\\1\end{array}\right)=1,  \nonumber\\
\vec{\al}&\cdot&\vec{V}_{S,\De,\ell}\succeq  0,   ~~~~~~~~~\De\geqslant\De_{S,0}^* ~~\text{for the $O(N)$ singlet scalars except $\sigma$, } \nonumber\\
\vec{\al}&\cdot&\vec{V}_{T,\De,\ell}\geqslant0,    \label{constraints}\\
\vec{\al}&\cdot&\vec{V}_{A,\De,\ell}\geqslant0,    \nonumber\\
\vec{\al}&\cdot&\vec{V}_{V,\De,\ell}\geqslant0,   ~~~~~~~~~\De\geqslant\De_{V,0}^* ~~\text{for the $O(N)$ vector scalars except $\phi_i$}, \nonumber \\
\vec{\al}&\cdot&\left(\vec{V}_{S,\De_\sigma,0}+\vec{V}_{V,\De_\phi,0}\otimes\left(\begin{array}{cc}
1 ~&~ 0 \\ 0 ~&~ 0\end{array}\right) \right)\succeq0, \nonumber
\eea
then the crossing relations (\ref{booteq}) can never be satisfied and initial assumption on the spectra $(\De_\phi,\De_\sigma)$ have to be abandoned as unphysical.
In the bootstrap conditions (\ref{constraints}), we have required the $O(N)$ singlet scalars (except $\sigma$) have conformal dimensions
above a lower bound $\De^*_{S,0}$, and similarly a lower bound for $\De_{V,0}^*$ for $O(N)$ vector scalars in addition to $\phi_i$.
Besides, we have implicitly assumed that all the extra operators accord with the unitary bound. In the last equation of (\ref{constraints}), it is the summation of
contributions in S (from $\sigma$) and V channels (from $\phi$) that is required to be positive-semidefinite due to the equality of OPE coefficients
$\la_{\phi\phi\sigma}=\la_{\phi\sigma\phi}$.

The bootstrap conditions in (\ref{constraints}) are not the only way to break the crossing relations (\ref{booteq}). In particular, to bootstrap certain
OPE coefficient of operator $(\De_0, \ell_0)$ in channel $X$, one may set $\vec{\al}\cdot\vec{V}_{X,\De_0,\ell_0}=1$ instead of choosing the unit operator as
in (\ref{constraints}). The bootstrap conditions are further refined in \cite{Kos:2016ysd}.
The lower bounds $\De_{S,0}^*$ and $\De_{V,0}^*$ introduced in (\ref{constraints}) are necessary to isolate the conformal dimensions $(\De_\phi, \De_\sigma)$
in a small island. A higher but remaining physical lower bound can improve the numerical efficiency to carve out the allowed parameter space. For sufficient
large $N$, these lower bounds can be justified from perturbative expansions. The $O(N)$ singlet scalar next to $\phi^2$ is
$\phi^4$ in the quartic model, and its conformal
dimension can be evaluated through the large $N$ expansion (\ref{Dsigma2}). In the cubic theory (\ref{lag}) this is given by a mixing of $\sigma^2$ and $\phi^2$.
One of the linear combination of $\sigma^2$ and $\phi^2$ is actually the descendent of $\sigma$, while another orthogonal mixing constructs a primary $O(N)$
singlet that shares the same conformal dimension as obtained from quartic theory \cite{Fei:2014yja, Fei:2014xta}.
The candidate of next $O(N)$ vector scalar is $\phi^2\phi_i$
(or $\sigma\phi_i$ in the cubic theory).
However, as argued in \cite{Kos:2015mba} for the $3D$ theories, in $D=6-\epsilon, ~\epsilon\ll1$ dimension the quartic theory generates the following equation of
motion for $\phi_i$:
\beq
\partial^2\phi_i\propto\phi^2\phi_i,
\eeq
which suggests that the operator $\phi^2\phi_i$ is a descendent of $\phi_i$ rather than a conformal primary scalar.
One can get the same conclusion in cubic theory (\ref{lag}) with replacement $\phi^2\rightarrow\sigma$.
The next candidate is $\phi^4\phi_i$ (in $D=6-\epsilon,~\epsilon\ll1$ dimension operators with derivatives,
like $\phi^2\partial^2\phi_i,(\partial_\mu \phi)^2\phi_i$ have different bare conformal dimensions given $\epsilon\neq0$ so they do not mix with $\phi^4\phi_i$).
At the interacting fixed point, the conformal dimension of $\phi^4\phi_i$ is expected to be studied using the large $N$ expansion.\footnote{Conformal primary
$O(N)$ vector scalars have been
studied in \cite{Lang:1992zw}; however, to our
knowledge, the explicit
perturbative result for $\phi^4\phi_i$ is still not available yet. We thank Simone Giombi for the valuable discussion on this problem.} At tree level the
conformal dimension of $\phi^2$ is $2$ near the interacting fixed point, so the conformal dimension of $\phi^4\phi_i$ is $5.5$ up to the order $O(1/N)$.
In the cubic theory the potential second $O(N)$ vector scalar is a mixing of $\sigma^2\phi_i$ and $\phi^2\phi_i$, which has not been explicitly studied yet.
One can expect that one of the mixing is actually a descendent of $\phi_i$ while another primary mixing has the same conformal dimension as $\phi^4\phi_i$
in quartic theory, as happened for the quadratic and cubic $O(N)$ singlet operators \cite{Fei:2014yja, Fei:2014xta}.
The lower bound of the $\phi^4\phi_i$ conformal dimension would be rather subtle for small $N$. Fortunately we will show that a unitary interacting fixed point
disappears even for $N=100$, indicating a large critical value $N_c$.

\subsection{Numerical Implementation of Conformal Bootstrap}
Equations from crossing symmetry (\ref{booteq}) provide an infinite set of constraints (\ref{constraints}) on the CFT data.
For the numerical implementation the constraints need to be truncated to a large but finite set. In (\ref{constraints}) the constraints are parameterized by
$(\De,\ell)$. The spins $\ell$ construct an infinite tower of spectra while in conformal bootstrap only these spectra with small $\ell$ will be considered. Contributions from operators with large spin are exponentially suppressed.
The linear functionals $\vec{\al}$ can be expanded as
\beq
\al_i=\sum_{m+n\leqslant\Lambda} a_{imn} \partial_z^m\partial_{\bar{z}}^n, \label{function}
\eeq
where $(z, \bar{z})$ are defined in terms of $(u, v)$ through: $u=z\bar{z}$, $v=(1-z)(1-\bar{z})$. Moreover, for the linear functional $\al_i$, the number of derivatives is also truncated up to $\Lambda$.
Taking higher order of derivatives in (\ref{function}), we have more chances to find the linear function satisfying (\ref{constraints}). As a result,
the conformal bootstrap program can exclude larger regions in parameter space. In practice the parameter $\Lambda$ is restricted by computation power. The setups of parameter $\Lambda$ and spins used in this work are as follows
\bea
S_{\Lambda=19}&=&\{0,1,\cdots,30\}\cup\{49,50\}, \nonumber\\
S_{\Lambda=21} &=& \{0,1,\cdots,30\}\cup\{47,48,49,50,51,52\}, \nonumber\\
S_{\Lambda=23} &=& \{0,1,\cdots,30\}\cup\{47,48,49,50,51,52,53,54\}, \nonumber\\
S_{\Lambda=25}&=&\{0,1,\cdots,30\}\cup\{47,48,49,50,51,52,53,54,55,56\}. \label{spins}
\eea
The problem to find the linear functions $\vec{\al}$ under truncated constraints can be solved with SDPB program \cite{Simmons-Duffin:2015qma}.

\section{Results}
\subsection{Bootstrapping $5D$ $O(500)$ Vector Model}
The $5D$ $O(500)$ vector model has been studied in \cite{Nakayama:2014yia, Bae:2014hia, Chester:2014gqa} using conformal bootstrap with single
correlator $\langle\phi_i\phi_j\phi_k\phi_l\rangle$. At the fixed point the conformal dimensions $(\De_\phi,\De_\sigma)$ of the lowest $O(N)$ vector $\phi_i$ and $O(N)$ singlet $\sigma$
can be evaluated from
the large $N$ expansion in (\ref{Dphi}, \ref{Dsigma}) or the $\epsilon$ expansion in (\ref{Dphie}, \ref{Dsigmae}).
Taking $N=500$, we get $(\De_\phi,\De_\sigma)=(1.500414, 2.02158)$ from 3-loop large $N$ expansion and  $(\De_\phi,\De_\sigma)=(1.500400, 2.02156)$ from
5-loop $\epsilon$ expansion. These predictions will be compared with the results obtained from conformal bootstrap.

\begin{figure}
\includegraphics[scale=0.9]{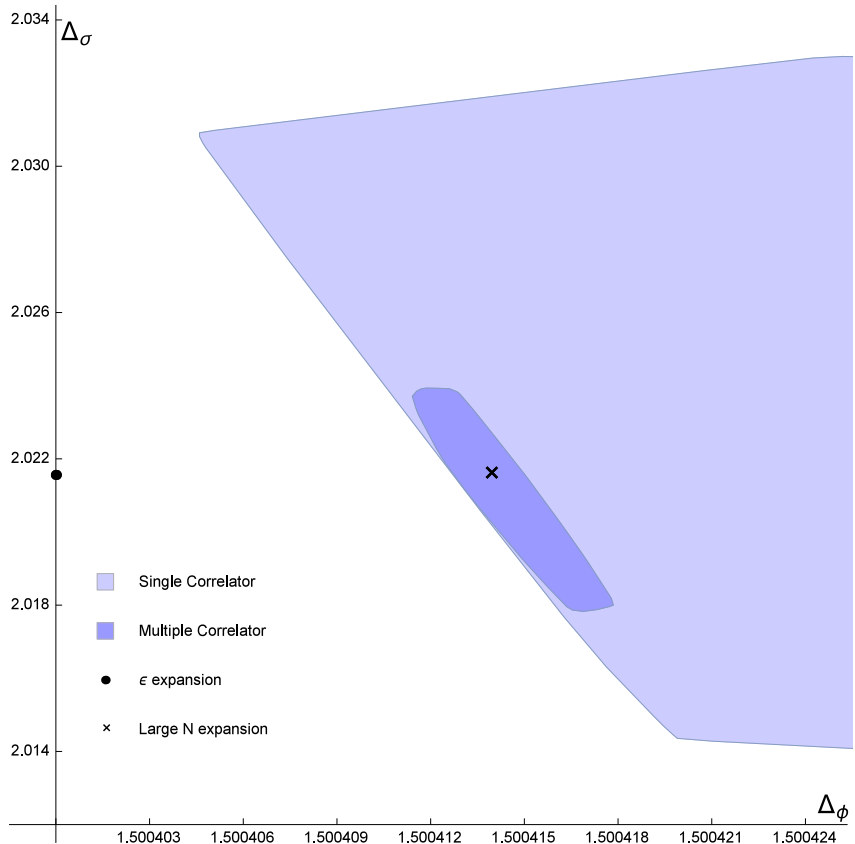}
 \begin{flushright}
\caption{Bounds on the conformal dimensions $(\De_\phi,\De_\sigma)$ in the interacting $5D$ $O(500)$ CFT. The colored regions represent the conformal dimensions allowed by conformal bootstrap. Specifically the light blue region is obtained from single correlator bootstrap, while the dark blue island is isolated
 through bootstrapping the multiple correlators. We used the derivative at order $\Lambda=19$ and spins $S_{\Lambda=19}$ in the numerical calculations. Besides, we assumed
a gap $\De_{S,0}^*=3.965$ in the S-channel. An extra gap $\De_{V,0}^*=5$ has been used in the V-channel for bootstrapping multiple correlators. The black dot and cross relate to the predictions from $\epsilon$ expansion and large $N$ expansion, respectively.} \label{N500}
\end{flushright}
\end{figure}

In Figure \ref{N500} we present the  bounds on $(\De_\phi,\De_\sigma)$ obtained through bootstrapping the single correlator $\langle\phi_i\phi_j\phi_k\phi_l\rangle$ (light blue region) and
the multiple correlators (dark blue island). To bootstrap the single correlator we have assumed that the next
$O(N)$ singlet scalar has dimension above the gap $\De_{S,0}^*=3.965$, which can be justified from the large $N$ expansion result ($\ref{Dsigma2}$):
$\De_{\sigma^2}\approx3.972$.
 This gap is also employed in \cite{Chester:2014gqa}.  The upper part of light blue region is similar to the bound provided in  \cite{Chester:2014gqa}. Besides, there is
an extra kink in the lower region and the whole region actually forms a sharp tip like presented in \cite{Bae:2014hia}, although a much larger gap was used
in that work. Results of perturbative methods are also shown in Figure 1. Prediction from the large $N$ expansion (denoted by the black cross) lies in the allowed region while
prediction from the $\epsilon$ expansion (denoted by the black dot) is outside of the bound so is excluded. According to the conformal bootstrap results, the large $N$ expansion does provide
a better estimation on the conformal dimensions for large $N=500$. Difference between the two perturbative approaches appears at the order $10^{-5}\approx O(1/N^2)$, as discussed before.

\begin{figure}
\includegraphics[scale=0.9]{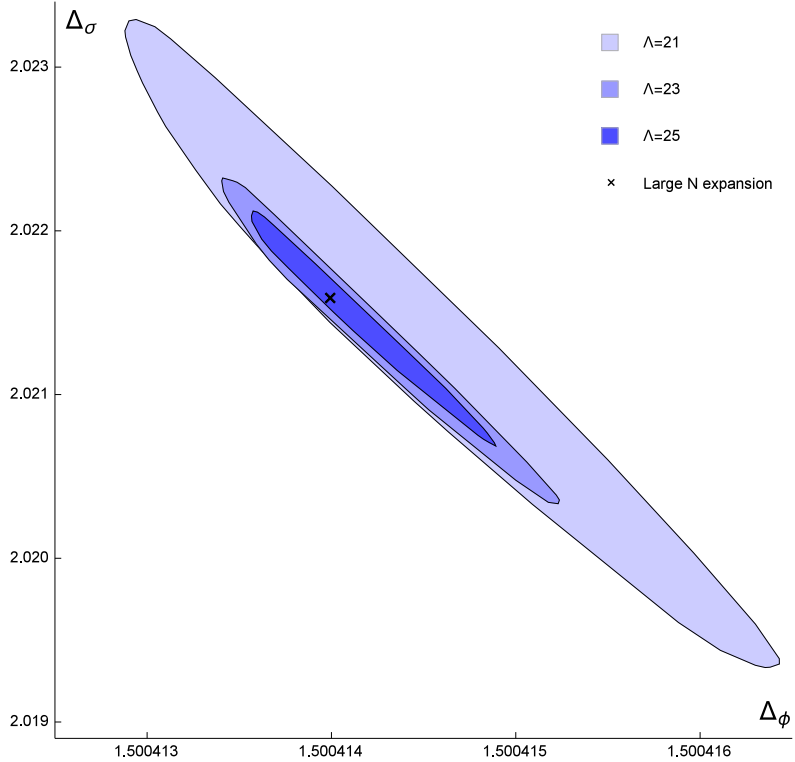}
 \begin{flushright}
\caption{Isolated regions for the conformal dimensions $(\De_\phi,\De_\sigma)$ in $5D$ $O(500)$ vector model. The light, medium and dark blue regions are corresponding to the results from multiple correlator conformal bootstrap with $\Lambda=21, ~23, ~25$, respectively. In the graph we have used the dimension gaps $\De_{S,0}^*=3.965$ and $\De_{V,0}^*=5$. The black cross denotes the prediction from large $N$ expansion.} \label{N500d}
\end{flushright}
\end{figure}

Remarkably, the allowed region of $(\De_\phi,\De_\sigma)$ obtained from the multiple correlator bootstrap is enclosed in a small island, which is colored
in dark blue in Figure 1. Besides the dimension gap $\De_{S,0}^*=3.965$ in S-channel, we have employed another dimension gap $\De_{V,0}^*=5$ in V-channel that the next primary $O(N)$ vector scalar has dimension $\De\geqslant5$.
The dark blue island lies in the center of the tip, and the black cross denoting the large $N$ prediction is rather close to the center of this island.
Such a high coincidence is extraordinary in view of only crossing symmetry and unitary condition are applied to carve out the island.
On the other hand, the conformal bootstrap result also shows that the large $N$ expansion is reliable at third order.\footnote{Strictly speaking, such consistency
check is not completely self-contained since we have already used the large $N$ expansion result in setting the dimension gaps.}

However, it should be careful to make statement based on results from conformal bootstrap with lower order of derivatives.
Actually in preliminary study we have obtained isolated islands even for $N=1$ with $\Lambda\sim15$; however, they disappear after increasing $\Lambda$. As to the model with $N=500$,
we have checked the performance of the island with larger $\Lambda$. The results are provided in Figure \ref{N500d}. The allowed regions shrink notably from $\Lambda=21$ to $\Lambda=25$. Interestingly, the fixed point predicted by large $N$ expansion remains located in the center of the small island even though the allowed region has contracted significantly.

\subsection{Bootstrapping $5D$ $O(N)$ $(N\leqslant100)$ Vector Models and the Critical $N_c$}
In $5D$ there is an interesting problem on the unitarity of the interacting $O(N)$ CFTs, that there is a threshold value $N_c$ below which the CFTs become nonunitary \cite{Fei:2014yja, Fei:2014xta}. In contrast, the interacting $O(N)$ CFTs in $3D$ are unitary for any integer $N\geqslant1$. Prior to our work, there are several evidences from conformal bootstrap which prefer to small $N_c$ \cite{Nakayama:2014yia, Bae:2014hia, Chester:2014gqa}. There are also some clues from perturbative results that the critical value $N_c<100$.
In this part we apply the conformal bootstrap with multiple correlators to study the $5D$ $O(N)$ vector model for small $N$s. The multiple correlator conformal bootstrap involves in more $O(N)$ sectors and provides stronger constraints on the CFT data comparing with the conformal bootstrap with single correlator only.

\begin{figure}
\includegraphics[scale=0.65]{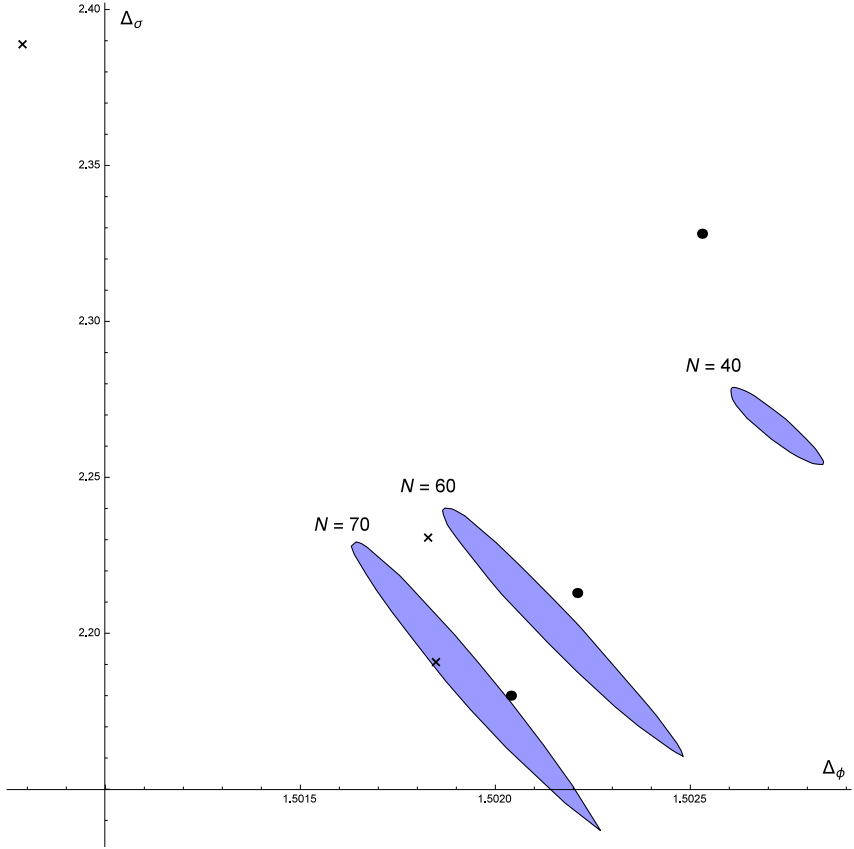}
 \begin{flushright}
\caption{From top to bottom, the islands represent the allowed regions of $(\De_\phi,\De_\sigma)$ in the $5D$ $O(N)$ $N=40,60,70$ vector models. The results are obtained from conformal bootstrap with $\Lambda=19$ and spins $S_{\Lambda=19}$. The black dots and crosses denote predictions from $\epsilon$ expansion and large $N$ expansions, respectively. The dimension gaps used in conformal bootstrap program are: $(\De_{S,0}^*,\De_{V,0}^*)=(3.4, 4.1)$ for $N=40$, $(\De_{S,0}^*,\De_{V,0}^*)=(3.5, 4.3)$ for $N=60, 70$.
The perturbative methods, especially the large $N$ expansion get abnormal and stay away from the region allowed by conformal bootstrap at $N=40$.} \label{3islands}
\end{flushright}
\end{figure}

We have searched the allowed regions on $(\De_\phi,\De_\sigma)$ plane for $N\leqslant100$. The isolated islands can be obtained for small $N$s with mild assumptions on the dimension gaps $(\De_{S,0}^*, \De_{V,0}^*)$. However, these islands disappear after increasing the number of derivatives $\Lambda$. For $N\sim O(10)$ or smaller, the perturbative approaches cannot provide an approximate estimation on the conformal dimension $\De_{\sigma^2}$. One may argue that the islands disappear due to the reason of the unphysical dimension gaps $(\De_{S,0}^*$, $\De_{V,0}^*)$ used in the bootstrap program instead of the nonunitarity of the CFTs. While for larger $N$s the perturbative predictions are expected to provide rough estimations on the fixed point. This can be seen from the fact that the isolated islands obtained from conformal bootstrap are close to the perturbative predictions before vanishing. In Figure \ref{3islands} we provide the isolated regions for
$N=40, 60, 70$ from conformal bootstrap. At derivative order $\Lambda=19$, the conformal bootstrap program generates closed regions on the $(\De_\phi,\De_\sigma)$ plane, which disappear for larger $\Lambda\geqslant23$. According to the results from conformal bootstrap, for $N=60, 70$ the perturbative approaches can still provide approximate estimations on the conformal dimensions at the interacting fixed points, although the theories are not unitary. While for $N=40$, the perturbative approaches, especially the large N expansion cannot provide reliable estimations on the interacting fixed point. One may note that the island corresponding to $N=40$ shown in Figure \ref{3islands} is rather close to the kink from single correlator bootstrap presented in \cite{Chester:2014gqa}. In \cite{Chester:2014gqa} the kink was considered to indicate a unitary CFT. However, our studies based on multiple correlator bootstrap show that the theory is actually not unitary and the kink, or the island before its vanishing uncovers an interacting while nonunitary CFT.

\begin{figure}
\includegraphics[scale=0.9]{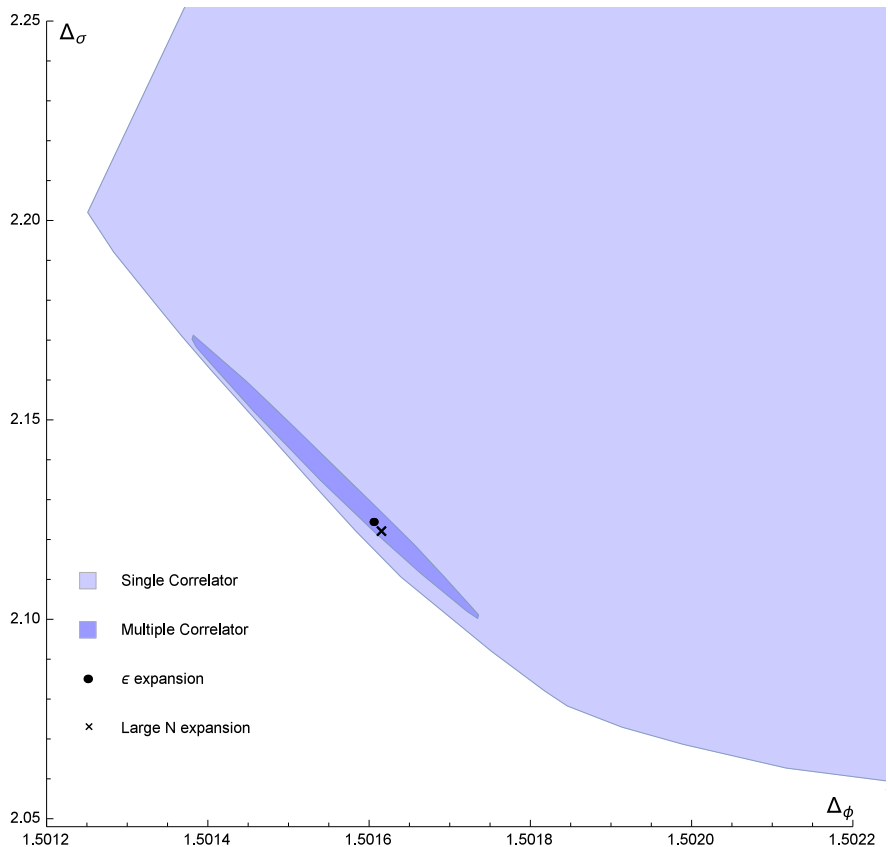}
 \begin{flushright}
\caption{
Bounds on the conformal dimensions $(\De_\phi,\De_\sigma)$ in $5D$ $O(100)$ vector model. The light blue region is obtained from single correlator bootstrap. The multiple correlators bootstrap leads to a small island colored in dark blue. In the bootstrap program we adopt the setup with $\Lambda=19$ and the correspond spins provided in (\ref{spins}). We apply a dimension gap $\De_{S,0}^*=3.6$ in the S-channel. Besides, an extra dimension gap $\De_{V,0}^*=5$ has been used in the V-channel for bootstrapping multiple correlators. The black dot and cross relate to the predictions from $\epsilon$  expansion and large $N$ expansion, respectively.} \label{N100ms}
\end{flushright}
\end{figure}

In fact there is no stable island from conformal bootstrap even at $N=100$. The perturbative methods predict that the interacting $O(100)$ fixed point locates in the position with conformal dimensions
$(\De_\phi,\De_\sigma)=(1.50161,2.124)$ from large $N$ expansion and $(\De_\phi,\De_\sigma)=(1.50162,2.122)$ from $\epsilon$ expansion.
In Figure \ref{N100ms} we show the conformal bootstrap results of $O(100)$ vector model with $\Lambda=19$. The single correlator conformal bootstrap generates a kinked bound similar to that of $O(500)$ vector model. The isolated region from multiple correlator conformal bootstrap lies in the middle of the tip. Here we have assumed a dimension gap $\De_{S,0}^*=3.6$ in the S-channel, which is considerably lower than the large $N$ prediction $\De_{\sigma^2}\approx3.850$. Besides, in the V-channel a dimension gap $\De_{V,0}^*=5$ has been used. Predictions from large $N$ and $\epsilon$ expansions are presented in the graph, both of which are nicely consistent with the conformal bootstrap bounds. In particular they locate in the isolated small island. All these features indicate a promising fixed point satisfying the crossing symmetry and unitarity constraints. However, the island disappears by taking higher order of derivatives $\Lambda=23$! Unless the ``true" island shrinks so drastically from $\Lambda=21$ to $\Lambda=23$ that it is hardly to be detected by scanning the parameter space, our bootstrap results disprove a unitary $5D$
$O(N)$ vector model even with $N=100$!

Vanishing of the ``allowed region" for $N\leqslant100$ suggests that the theory actually is not unitary.  The violation of unitarity is rather small so that it cannot be uncovered by the bootstrap program with smaller $\Lambda$. This reminds us other
examples on ``pseudo" unitarity in conformal bootstrap. In \cite{El-Showk:2013nia} the $O(N)$ vector models in fractional dimensions $2<D<4$ have been studied
using conformal bootstrap. In the work pronounced kinks are obtained in the bounds of conformal dimension of the lowest $O(N)$ singlet $\sigma$ and are well consistent with the
results obtained from extra approaches. However, careful studies in \cite{Hogervorst:2014rta, Hogervorst:2015akt} have shown that the CFTs in fractional dimensions are
necessarily to be nonunitary, which are too subtle to be discovered in numerical conformal bootstrap. In the $5D$ $O(N)$ single correlator
conformal bootstrap \cite{Chester:2014gqa},
sharp kinks are also generated in the fractional dimension $D=5.95$ with $N=600$, notably lower than the critical value $N_c\approx1000$. We have studied this model through bootstrapping multiple correlators. There remains isolated allowed region even at $\Lambda=21$, though it is quite small. The uncertainty on $\De_\sigma$ shown in the island is about $2\times10^{-3}$, while as shown in \cite{Chester:2014gqa}, the magnitude of imaginary part in $\De_\sigma$ is of the
same order $\sim1.5\times10^{-3}$ so it is expected that current conformal bootstrap program cannot capture the tiny unitarity violation unless the numerical accuracy can be improved significantly.

To summarize, the numerical conformal bootstrap provides a powerful approach to falsify assumptions on unitary CFTs. However, it is premature to validate the unitary CFTs using conformal bootstrap due to these ``pseudo" unitary solutions. As to the $5D$ $O(500)$ model, although our results have provided strong evidence, they are still not sufficient to make a strict conclusion on its unitarity. On the other hand, it is surprising that the $5D$ $O(N)$ vector model is nonunitary even for $N=100$. As a result, the critical value $N_c>100$, which is
considerably larger than the value estimated before.
\section{Conclusions}
In this work, we have studied the interacting $5D$ CFTs with global $O(N)$ symmetry using the conformal bootstrap with multiple correlators. The multiple correlator conformal bootstrap has been developed in \cite{Kos:2014bka, Kos:2015mba} and obtained remarkable successes in $3D$ Ising and $O(N)$ vector models. The approach employs the correlators of the $O(N)$ vector scalar $\phi_i$ as well as the $O(N)$ singlet scalar $\sigma$. Since there are more operators involved
in the crossing symmetry relations,
the new method is expected to generate more strong constraints on the CFT data. Indeed the allowed regions on ($\De_\phi,\De_\sigma$) plane is limited in a small island under reasonable assumptions on the dimension gaps.

Specifically, in this study we have shown that by bootstrapping multiple correlators from the interacting $5D$ CFTs with $O(N)$  symmetry ($N=500$), the allowed conformal
dimensions ($\De_\phi,\De_\sigma$)  are strongly limited in a closed region, which is highly consistent with predictions from large $N$ expansion. In order to uncover the isolated region we also applied assumptions on the dimension gaps both in the
$O(N)$ singlet sector and the $O(N)$ vector sector.
Our results suggest that the interacting fixed point of $O(N)$ vector model is unitary for sufficient large $N$ and support the asymptotic free $5D$ $O(N)$ cubic model proposed in \cite{Fei:2014yja, Fei:2014xta}. Evidence of such fixed point has already been shown in the single correlator conformal bootstrap
studied in \cite{Nakayama:2014yia, Bae:2014hia, Chester:2014gqa}. The island obtained in this work is rather close to the kink in the bound of conformal dimension $\De_\sigma$ obtained from bootstrapping correlator of four $\phi_i$s \cite{Chester:2014gqa}.
We have studied the performance of the island under higher order of derivatives $\Lambda$. The island shrinks notably from $\Lambda=19$ to $\Lambda=25$, while the large $N$ expansion predictions remain staying in the center of the allowed region. Such coincidence is surprising in considering of that only crossing symmetry and unitary conditions are employed to generate the allowed region. Besides we only input the $O(N)$ global symmetry for this model while even did not use its Lagrangian at all.

We are particularly interested in the critical value $N_c$ of $5D$ $O(N)$ vector model below which the interacting fixed point theory loses unitarity.
The problem on the critical value $N_c$ can also be seen from the perturbative expansions of conformal dimension $\De_\phi$, that below the critical value the
scalar $\phi_i$ acquires conformal dimension smaller than the unitary bound and breaks the unitary condition. However, in $5D$ the perturbative expansions converges much slower comparing with these of $3D$.
In \cite{Fei:2014yja, Fei:2014xta} the critical value $N_c$ has been evaluated based on large $N$ expansion in  $D=6-\epsilon$ spacetime. The critical value $N_c\simeq1038$ at one-loop level; however, it oscillates drastically order by order. Conformal bootstrap provides a nonperturbative approach to study CFTs, and it has been applied to estimate $N_c$ in \cite{Chester:2014gqa}. The authors found that the pronounced kink in the bound of $\De_\sigma$ disappears near $N\sim15$, which may suggest $N_c\sim15$ in view of the observation that the singular behaviors, like kink in the dimension bound usually relate to unitary CFTs. In $3D$ such observation has helped to numerically solve the Ising model \cite{El-Showk:2014dwa} and $O(N)$ vector model \cite{Kos:2013tga}.
However, the unitarity condition becomes subtle for $5D$ CFTs and the unitarity violation may be too small to be detected by the bootstrap program with low order of derivatives. Therefore a kink does not necessarily guarantee unitarity, instead, it may relate to
an interacting but nonunitary CFTs.

We have searched the allowed regions using multiple correlator conformal bootstrap for $1\leqslant N\leqslant100$. The isolated regions on the  $(\De_\phi,\De_\sigma)$ plan can be obtained from conformal bootstrap program with lower order of derivatives.  Moreover, the islands actually locate in the position close to the predictions from perturbative approaches given the $N$s are not too small. However, the islands
disappear after increasing the number of derivatives in bootstrap program. We believe these islands relate to interacting while nonunitary CFTs and the violation of unitarity can not be observed unless the program is equipped with sufficient high precision. In particular, our results suggest the critical value $N_c\geqslant100$, much larger than the value estimated before. The bounds of $N_c$ is expected to be improved further using conformal bootstrap. However, for larger $N$ the unitarity violation in $O(N)$ fixed point theory gets smaller and more difficult to be detected. It requires higher accuracy in the bootstrap program to determine the critical value $N_c$ and  we leave this problem for future work. On the other hand, for a sufficient large $N_c$, the large $N$ expansion approach is validated. The critical value $N_c$ can be effectively studied based on this perturbative approach as well. Due to the asymptotic
behavior of perturbative expansions in $5D$, probably one needs to calculate several orders higher than in \cite{Fei:2014yja, Fei:2014xta, Gracey:2015tta} to get a sufficient good estimation.

\section*{Acknowledgements}
We are particularly grateful to David Simmons-Duffin for helpful correspondences on the numerical conformal bootstrap.
We thank Simone Giombi for his comments on the spectra of $O(N)$ vector sector. We also thank Daliang Li, Tianjun Li, Daniel Robbins and Junchen Rong for valuable discussions. To calculate the conformal block functions, we used the code from an open-source program JuliBootS which is available at \url{https://github.com/mfpaulos/JuliBoots}. The work of N.S. was supported by NSF grants PHY-1214333 and PHY-1521099.

%%%%%%%%%%%%%
%%%%%%%%%%%%
%%%%%%%%%%%
%%%%%%%%%%
%%%%%%%%%%%
%%%%%%%%%%%%
%%%%%%%%%%%%%

\newpage
\appendix
\section{Bootstrap Equations}
In (\ref{booteq}) the crossing symmetry relations have been summarized in a compact form, as in \cite{Kos:2015mba} for $3D$ $O(N)$ vector model. The seven bootstrap equations obtained from $O(N)$ singlet ($S$), traceless symmetric tensor ($T$), antisymmetric tensor ($A$) and vector ($V$) sectors of multiple
correlators are summarized in a 7-component vector equation (\ref{booteq}),
in which the vectors $\vec{V}_S, \vec{V}_{T}, \vec{V}_A, \vec{V}_V$ are:
\bea
\vec{V}_{T,\De,\ell} = \left( \begin{array}{c} F_{-,\De,\ell}^{\phi\phi,\phi\phi}  \\ \left(1-\frac2N\right)F_{-,\De,\ell}^{\phi\phi,\phi\phi}
\\ -\left(1+\frac{2}{N}\right)F_{+,\De,\ell}^{\phi\phi,\phi\phi}\\ \mbox{\textbf{\large 0}}_{4\times1} \end{array} \right),
\
\vec{V}_{A,\De,\ell} = \left( \begin{array}{c} - F_{-,\De,\ell}^{\phi\phi,\phi\phi}  \\  F_{-,\De,\ell}^{\phi\phi,\phi\phi}
\\ - F_{+,\De,\ell}^{\phi\phi,\phi\phi}\\ \mbox{\textbf{\large 0}}_{4\times1} \end{array} \right),
\
\vec{V}_{V,\De,\ell} = \left( \begin{array}{c} \mbox{\textbf{\large 0}}_{4\times1} \\ (-1)^\ell F_{-,\De,\ell}^{\phi \sigma,\phi \sigma}\\
F_{-,\De,\ell}^{\sigma\phi,\phi \sigma}
\\- F_{+,\De,\ell}^{\sigma\phi,\phi \sigma}  \end{array} \right),\nonumber
\eea

\bea
\vec{V}_{S,\De,\ell} = \left( \begin{array}{c}
 \mbox{\textbf{\large 0}}_{2\times2} \\
\left( \begin{array}{cc} F^{\phi\phi,\phi\phi}_{-,\De,\ell}(u,v) & 0 \\ 0 & 0 \end{array} \right) \\
\left( \begin{array}{cc} F^{\phi\phi,\phi\phi}_{+,\De,\ell}(u,v) & 0 \\ 0 & 0 \end{array} \right) \\
 \left( \begin{array}{cc} 0 & 0 \\ 0 & F^{\sigma\sigma,\sigma\sigma}_{-,\De,\ell}(u,v) \end{array} \right) \\
\mbox{\textbf{\large 0}}_{2\times2} \\
 \left( \begin{array}{cc} 0 & \frac{1}{2} F^{\phi\phi,\sigma\sigma}_{-,\De,\ell}(u,v) \\ \frac{1}{2}
 F^{\phi\phi,\sigma\sigma}_{-,\De,\ell}(u,v) & 0 \end{array} \right) \\
 \left( \begin{array}{cc} 0 & \frac{1}{2} F^{\phi\phi,\sigma\sigma}_{+,\De,\ell}(u,v) \\ \frac{1}{2}
 F^{\phi\phi,\sigma\sigma}_{+,\De,\ell}(u,v) & 0 \end{array} \right) \end{array} \right).
\eea
Here our convention differs from \cite{Kos:2015mba} by a factor $(-1)^{\ell}$.


\begin{thebibliography}{99}
%\cite{Ferrara:1973yt, Polyakov:1974gs, Mack:1975jr}
\bibitem{Ferrara:1973yt}
  S.~Ferrara, A.~F.~Grillo and R.~Gatto,
  %``Tensor representations of conformal algebra and conformally covariant operator product expansion,''
  Annals Phys.\  {\bf 76}, 161 (1973).
  doi:10.1016/0003-4916(73)90446-6
%\cite{Polyakov:1974gs}
\bibitem{Polyakov:1974gs}
  A.~M.~Polyakov,
  %``Nonhamiltonian approach to conformal quantum field theory,''
  Zh.\ Eksp.\ Teor.\ Fiz.\  {\bf 66}, 23 (1974).
  %%CITATION = ZETFA,66,23;%%
%\cite{Mack:1975jr}
\bibitem{Mack:1975jr}
  G.~Mack,
  %``Duality in quantum field theory,''
  Nucl.\ Phys.\ B {\bf 118}, 445 (1977).
  doi:10.1016/0550-3213(77)90238-3
%\cite{Belavin:1984vu}
\bibitem{Belavin:1984vu}
  A.~A.~Belavin, A.~M.~Polyakov and A.~B.~Zamolodchikov,
  %``Infinite Conformal Symmetry in Two-Dimensional Quantum Field Theory,''
  Nucl.\ Phys.\ B {\bf 241}, 333 (1984).
  doi:10.1016/0550-3213(84)90052-X

%\cite{Rattazzi:2008pe}
\bibitem{Rattazzi:2008pe}
  R.~Rattazzi, V.~S.~Rychkov, E.~Tonni and A.~Vichi,
  %``Bounding scalar operator dimensions in 4D CFT,''
  JHEP {\bf 0812}, 031 (2008)
  doi:10.1088/1126-6708/2008/12/031
  [arXiv:0807.0004 [hep-th]].

%\cite{Dolan:2000ut, Dolan:2003hv}
\bibitem{Dolan:2000ut}
  F.~A.~Dolan and H.~Osborn,
  %``Conformal four point functions and the operator product expansion,''
  Nucl.\ Phys.\ B {\bf 599}, 459 (2001)
  doi:10.1016/S0550-3213(01)00013-X
  [hep-th/0011040].
%\cite{Dolan:2003hv}
\bibitem{Dolan:2003hv}
  F.~A.~Dolan and H.~Osborn,
  %``Conformal partial waves and the operator product expansion,''
  Nucl.\ Phys.\ B {\bf 678}, 491 (2004)
  doi:10.1016/j.nuclphysb.2003.11.016
  [hep-th/0309180].

%\cite{Rychkov:2009ij}
\bibitem{Rychkov:2009ij}
  V.~S.~Rychkov and A.~Vichi,
  %``Universal Constraints on Conformal Operator Dimensions,''
  Phys.\ Rev.\ D {\bf 80}, 045006 (2009)
  doi:10.1103/PhysRevD.80.045006
  [arXiv:0905.2211 [hep-th]].
%\cite{Caracciolo:2009bx}
\bibitem{Caracciolo:2009bx}
  F.~Caracciolo and V.~S.~Rychkov,
  %``Rigorous Limits on the Interaction Strength in Quantum Field Theory,''
  Phys.\ Rev.\ D {\bf 81}, 085037 (2010)
  doi:10.1103/PhysRevD.81.085037
  [arXiv:0912.2726 [hep-th]].
%\cite{Poland:2010wg}
\bibitem{Poland:2010wg}
  D.~Poland and D.~Simmons-Duffin,
  %``Bounds on 4D Conformal and Superconformal Field Theories,''
  JHEP {\bf 1105}, 017 (2011)
  doi:10.1007/JHEP05(2011)017
  [arXiv:1009.2087 [hep-th]].
%\cite{Rattazzi:2010gj}
\bibitem{Rattazzi:2010gj}
  R.~Rattazzi, S.~Rychkov and A.~Vichi,
  %``Central Charge Bounds in 4D Conformal Field Theory,''
  Phys.\ Rev.\ D {\bf 83}, 046011 (2011)
  doi:10.1103/PhysRevD.83.046011
  [arXiv:1009.2725 [hep-th]].
%\cite{Rattazzi:2010yc}
\bibitem{Rattazzi:2010yc}
  R.~Rattazzi, S.~Rychkov and A.~Vichi,
  %``Bounds in 4D Conformal Field Theories with Global Symmetry,''
  J.\ Phys.\ A {\bf 44}, 035402 (2011)
  doi:10.1088/1751-8113/44/3/035402
  [arXiv:1009.5985 [hep-th]].
%\cite{Vichi:2011ux}
\bibitem{Vichi:2011ux}
  A.~Vichi,
  %``Improved bounds for CFT's with global symmetries,''
  JHEP {\bf 1201}, 162 (2012)
  doi:10.1007/JHEP01(2012)162
  [arXiv:1106.4037 [hep-th]].
%\cite{Poland:2011ey}
\bibitem{Poland:2011ey}
  D.~Poland, D.~Simmons-Duffin and A.~Vichi,
  %``Carving Out the Space of 4D CFTs,''
  JHEP {\bf 1205}, 110 (2012)
  doi:10.1007/JHEP05(2012)110
  [arXiv:1109.5176 [hep-th]].
%\cite{ElShowk:2012ht}
\bibitem{ElShowk:2012ht}
  S.~El-Showk, M.~F.~Paulos, D.~Poland, S.~Rychkov, D.~Simmons-Duffin and A.~Vichi,
  %``Solving the 3D Ising Model with the Conformal Bootstrap,''
  Phys.\ Rev.\ D {\bf 86}, 025022 (2012)
  doi:10.1103/PhysRevD.86.025022
  [arXiv:1203.6064 [hep-th]].
%\cite{Liendo:2012hy}
\bibitem{Liendo:2012hy}
  P.~Liendo, L.~Rastelli and B.~C.~van Rees,
  %``The Bootstrap Program for Boundary CFT_d,''
  JHEP {\bf 1307}, 113 (2013)
  doi:10.1007/JHEP07(2013)113
  [arXiv:1210.4258 [hep-th]].
%\cite{Beem:2013qxa}
\bibitem{Beem:2013qxa}
  C.~Beem, L.~Rastelli and B.~C.~van Rees,
  %``The $\mathcal N=4$ Superconformal Bootstrap,''
  Phys.\ Rev.\ Lett.\  {\bf 111}, 071601 (2013)
  doi:10.1103/PhysRevLett.111.071601
  [arXiv:1304.1803 [hep-th]].
%\cite{Kos:2013tga}
\bibitem{Kos:2013tga}
  F.~Kos, D.~Poland and D.~Simmons-Duffin,
  %``Bootstrapping the $O(N)$ vector models,''
  JHEP {\bf 1406}, 091 (2014)
  doi:10.1007/JHEP06(2014)091
  [arXiv:1307.6856 [hep-th]].
%\cite{El-Showk:2013nia}
\bibitem{El-Showk:2013nia}
  S.~El-Showk, M.~Paulos, D.~Poland, S.~Rychkov, D.~Simmons-Duffin and A.~Vichi,
  %``Conformal Field Theories in Fractional Dimensions,''
  Phys.\ Rev.\ Lett.\  {\bf 112}, 141601 (2014)
  doi:10.1103/PhysRevLett.112.141601
  [arXiv:1309.5089 [hep-th]].
%\cite{Alday:2013opa}
\bibitem{Alday:2013opa}
  L.~F.~Alday and A.~Bissi,
  %``The superconformal bootstrap for structure constants,''
  JHEP {\bf 1409}, 144 (2014)
  doi:10.1007/JHEP09(2014)144
  [arXiv:1310.3757 [hep-th]].
%\cite{Gaiotto:2013nva}
\bibitem{Gaiotto:2013nva}
  D.~Gaiotto, D.~Mazac and M.~F.~Paulos,
  %``Bootstrapping the 3d Ising twist defect,''
  JHEP {\bf 1403}, 100 (2014)
  doi:10.1007/JHEP03(2014)100
  [arXiv:1310.5078 [hep-th]].
%\cite{Berkooz:2014yda}
\bibitem{Berkooz:2014yda}
  M.~Berkooz, R.~Yacoby and A.~Zait,
  %``Bounds on $\mathcal{N} = 1$ superconformal theories with global symmetries,''
  JHEP {\bf 1408}, 008 (2014)
  Erratum: [JHEP {\bf 1501}, 132 (2015)]
  doi:10.1007/JHEP01(2015)132, 10.1007/JHEP08(2014)008
  [arXiv:1402.6068 [hep-th]].
%\cite{El-Showk:2014dwa}
\bibitem{El-Showk:2014dwa}
  S.~El-Showk, M.~F.~Paulos, D.~Poland, S.~Rychkov, D.~Simmons-Duffin and A.~Vichi,
  %``Solving the 3d Ising Model with the Conformal Bootstrap II. c-Minimization and Precise Critical Exponents,''
  J.\ Stat.\ Phys.\  {\bf 157}, 869 (2014)
  doi:10.1007/s10955-014-1042-7
  [arXiv:1403.4545 [hep-th]].
%\cite{Nakayama:2014lva}
\bibitem{Nakayama:2014lva}
  Y.~Nakayama and T.~Ohtsuki,
  %``Approaching the conformal window of $O(n)\times O(m)$ symmetric Landau-Ginzburg models using the conformal bootstrap,''
  Phys.\ Rev.\ D {\bf 89}, no. 12, 126009 (2014)
  doi:10.1103/PhysRevD.89.126009
  [arXiv:1404.0489 [hep-th]].
%\cite{Nakayama:2014yia}
\bibitem{Nakayama:2014yia}
  Y.~Nakayama and T.~Ohtsuki,
  %``Five dimensional $O(N)$-symmetric CFTs from conformal bootstrap,''
  Phys.\ Lett.\ B {\bf 734}, 193 (2014)
  doi:10.1016/j.physletb.2014.05.058
  [arXiv:1404.5201 [hep-th]].
%\cite{Alday:2014qfa}
\bibitem{Alday:2014qfa}
  L.~F.~Alday and A.~Bissi,
  %``Generalized bootstrap equations for $ \mathcal{N}=4 $ SCFT,''
  JHEP {\bf 1502}, 101 (2015)
  doi:10.1007/JHEP02(2015)101
  [arXiv:1404.5864 [hep-th]].
%\cite{Chester:2014fya}
\bibitem{Chester:2014fya}
  S.~M.~Chester, J.~Lee, S.~S.~Pufu and R.~Yacoby,
  %``The $ \mathcal{N}=8 $ superconformal bootstrap in three dimensions,''
  JHEP {\bf 1409}, 143 (2014)
  doi:10.1007/JHEP09(2014)143
  [arXiv:1406.4814 [hep-th]].
%\cite{Kos:2014bka}
\bibitem{Kos:2014bka}
  F.~Kos, D.~Poland and D.~Simmons-Duffin,
  %``Bootstrapping Mixed Correlators in the 3D Ising Model,''
  JHEP {\bf 1411}, 109 (2014)
  doi:10.1007/JHEP11(2014)109
  [arXiv:1406.4858 [hep-th]].
%\cite{Caracciolo:2014cxa}
\bibitem{Caracciolo:2014cxa}
  F.~Caracciolo, A.~C.~Echeverri, B.~von Harling and M.~Serone,
  %``Bounds on OPE Coefficients in 4D Conformal Field Theories,''
  JHEP {\bf 1410}, 20 (2014)
  doi:10.1007/JHEP10(2014)020
  [arXiv:1406.7845 [hep-th]].
%\cite{Nakayama:2014sba}
\bibitem{Nakayama:2014sba}
  Y.~Nakayama and T.~Ohtsuki,
  %``Bootstrapping phase transitions in QCD and frustrated spin systems,''
  Phys.\ Rev.\ D {\bf 91}, no. 2, 021901 (2015)
  doi:10.1103/PhysRevD.91.021901
  [arXiv:1407.6195 [hep-th]].
%\cite{Golden:2014oqa}
\bibitem{Golden:2014oqa}
  J.~Golden and M.~F.~Paulos,
  %``No unitary bootstrap for the fractal Ising model,''
  JHEP {\bf 1503}, 167 (2015)
  doi:10.1007/JHEP03(2015)167
  [arXiv:1411.7932 [hep-th]].
%\cite{Chester:2014mea}
\bibitem{Chester:2014mea}
  S.~M.~Chester, J.~Lee, S.~S.~Pufu and R.~Yacoby,
  %``Exact Correlators of BPS Operators from the 3d Superconformal Bootstrap,''
  JHEP {\bf 1503}, 130 (2015)
  doi:10.1007/JHEP03(2015)130
  [arXiv:1412.0334 [hep-th]].
%\cite{Paulos:2014vya}
\bibitem{Paulos:2014vya}
  M.~F.~Paulos,
  %``JuliBootS: a hands-on guide to the conformal bootstrap,''
  arXiv:1412.4127 [hep-th].
  %%CITATION = ARXIV:1412.4127;%%
%\cite{Beem:2014zpa}
\bibitem{Beem:2014zpa}
  C.~Beem, M.~Lemos, P.~Liendo, L.~Rastelli and B.~C.~van Rees,
  %``The $ \mathcal{N}=2 $ superconformal bootstrap,''
  JHEP {\bf 1603}, 183 (2016)
  doi:10.1007/JHEP03(2016)183
  [arXiv:1412.7541 [hep-th]].
%\cite{Bae:2014hia, Chester:2014gqa}
\bibitem{Bae:2014hia}
  J.~B.~Bae and S.~J.~Rey,
  %``Conformal Bootstrap Approach to O(N) Fixed Points in Five Dimensions,''
  arXiv:1412.6549 [hep-th].
  %%CITATION = ARXIV:1412.6549;%%
%\cite{Chester:2014gqa}
\bibitem{Chester:2014gqa}
  S.~M.~Chester, S.~S.~Pufu and R.~Yacoby,
  %``Bootstrapping $O(N)$ vector models in 4 $< d <$ 6,''
  Phys.\ Rev.\ D {\bf 91}, no. 8, 086014 (2015)
  doi:10.1103/PhysRevD.91.086014
  [arXiv:1412.7746 [hep-th]].
%\cite{Simmons-Duffin:2015qma}
\bibitem{Simmons-Duffin:2015qma}
  D.~Simmons-Duffin,
  %``A Semidefinite Program Solver for the Conformal Bootstrap,''
  JHEP {\bf 1506}, 174 (2015)
  doi:10.1007/JHEP06(2015)174
  [arXiv:1502.02033 [hep-th]].
%\cite{Bobev:2015vsa}
\bibitem{Bobev:2015vsa}
  N.~Bobev, S.~El-Showk, D.~Mazac and M.~F.~Paulos,
  %``Bootstrapping the Three-Dimensional Supersymmetric Ising Model,''
  Phys.\ Rev.\ Lett.\  {\bf 115}, no. 5, 051601 (2015)
  doi:10.1103/PhysRevLett.115.051601
  [arXiv:1502.04124 [hep-th]].
%\cite{Kos:2015mba}
\bibitem{Kos:2015mba}
  F.~Kos, D.~Poland, D.~Simmons-Duffin and A.~Vichi,
  %``Bootstrapping the O(N) Archipelago,''
  JHEP {\bf 1511}, 106 (2015)
  doi:10.1007/JHEP11(2015)106
  [arXiv:1504.07997 [hep-th]].
%\cite{Chester:2015qca}
\bibitem{Chester:2015qca}
  S.~M.~Chester, S.~Giombi, L.~V.~Iliesiu, I.~R.~Klebanov, S.~S.~Pufu and R.~Yacoby,
  %``Accidental Symmetries and the Conformal Bootstrap,''
  JHEP {\bf 1601}, 110 (2016)
  doi:10.1007/JHEP01(2016)110
  [arXiv:1507.04424 [hep-th]].
%\cite{Beem:2015aoa}
\bibitem{Beem:2015aoa}
  C.~Beem, M.~Lemos, L.~Rastelli and B.~C.~van Rees,
  %``The (2, 0) superconformal bootstrap,''
  Phys.\ Rev.\ D {\bf 93}, no. 2, 025016 (2016)
  doi:10.1103/PhysRevD.93.025016
  [arXiv:1507.05637 [hep-th]].
%\cite{Iliesiu:2015qra}
\bibitem{Iliesiu:2015qra}
  L.~Iliesiu, F.~Kos, D.~Poland, S.~S.~Pufu, D.~Simmons-Duffin and R.~Yacoby,
  %``Bootstrapping 3D Fermions,''
  JHEP {\bf 1603}, 120 (2016)
  doi:10.1007/JHEP03(2016)120
  [arXiv:1508.00012 [hep-th]].
%\cite{Poland:2015mta}
\bibitem{Poland:2015mta}
  D.~Poland and A.~Stergiou,
  %``Exploring the Minimal 4D $\mathcal{N}=1$ SCFT,''
  JHEP {\bf 1512}, 121 (2015)
  doi:10.1007/JHEP12(2015)121
  [arXiv:1509.06368 [hep-th]].
%\cite{Lemos:2015awa}
\bibitem{Lemos:2015awa}
  M.~Lemos and P.~Liendo,
  %``Bootstrapping $ \mathcal{N}=2 $ chiral correlators,''
  JHEP {\bf 1601}, 025 (2016)
  doi:10.1007/JHEP01(2016)025
  [arXiv:1510.03866 [hep-th]].
%\cite{Lin:2015wcg}
\bibitem{Lin:2015wcg}
  Y.~H.~Lin, S.~H.~Shao, D.~Simmons-Duffin, Y.~Wang and X.~Yin,
  %``N=4 Superconformal Bootstrap of the K3 CFT,''
  arXiv:1511.04065 [hep-th].
%\cite{Chester:2015lej}
\bibitem{Chester:2015lej}
  S.~M.~Chester, L.~V.~Iliesiu, S.~S.~Pufu and R.~Yacoby,
  %``Bootstrapping $O(N)$ Vector Models with Four Supercharges in $3 \leq d \leq4$,''
  arXiv:1511.07552 [hep-th].
  %%CITATION = ARXIV:1511.07552;%%
%\cite{Chester:2016wrc}
\bibitem{Chester:2016wrc}
  S.~M.~Chester and S.~S.~Pufu,
  %``Towards Bootstrapping QED$_3$,''
  arXiv:1601.03476 [hep-th].
%\cite{Iha:2016ppj}
\bibitem{Iha:2016ppj}
  H.~Iha, H.~Makino and H.~Suzuki,
  %``Upper bound on the mass anomalous dimension in many-flavor gauge theories: a conformal bootstrap approach,''
  PTEP {\bf 2016}, no. 5, 053B03 (2016)
  doi:10.1093/ptep/ptw046
  [arXiv:1603.01995 [hep-th]].
%\cite{Kos:2016ysd}
\bibitem{Kos:2016ysd}
  F.~Kos, D.~Poland, D.~Simmons-Duffin and A.~Vichi,
  %``Precision Islands in the Ising and $O(N)$ Models,''
  arXiv:1603.04436 [hep-th].
  %%CITATION = ARXIV:1603.04436;%%

%\cite{Simmons-Duffin:2016gjk}
\bibitem{Simmons-Duffin:2016gjk}
  D.~Simmons-Duffin,
  %``TASI Lectures on the Conformal Bootstrap,''
  arXiv:1602.07982 [hep-th].
  %%CITATION = ARXIV:1602.07982;%%

%\cite{Pelissetto:2000ek}
\bibitem{Pelissetto:2000ek}
  A.~Pelissetto and E.~Vicari,
  %``Critical phenomena and renormalization group theory,''
  Phys.\ Rept.\  {\bf 368}, 549 (2002)
  doi:10.1016/S0370-1573(02)00219-3
  [cond-mat/0012164].

%\cite{Klebanov:2002ja}
\bibitem{Klebanov:2002ja}
  I.~R.~Klebanov and A.~M.~Polyakov,
  %``AdS dual of the critical O(N) vector model,''
  Phys.\ Lett.\ B {\bf 550}, 213 (2002)
  doi:10.1016/S0370-2693(02)02980-5
  [hep-th/0210114].

%\cite{Parisi:1975im}
\bibitem{Parisi:1975im}
  G.~Parisi,
  %``The Theory of Nonrenormalizable Interactions. 1. The Large N Expansion,''
  Nucl.\ Phys.\ B {\bf 100}, 368 (1975).
  doi:10.1016/0550-3213(75)90624-0
  %%CITATION = doi:10.1016/0550-3213(75)90624-0;%%
%\cite{Parisi:1977uz}
\bibitem{Parisi:1977uz}
  G.~Parisi,
  %``On Nonrenormalizable Interactions,''
  PRINT-77-0054 (IHES,BURES).

%\cite{Fei:2014yja, Fei:2014xta}
\bibitem{Fei:2014yja}
  L.~Fei, S.~Giombi and I.~R.~Klebanov,
  %``Critical $O(N)$ models in $6-\epsilon$ dimensions,''
  Phys.\ Rev.\ D {\bf 90}, no. 2, 025018 (2014)
  doi:10.1103/PhysRevD.90.025018
  [arXiv:1404.1094 [hep-th]].
%\cite{Fei:2014xta}
\bibitem{Fei:2014xta}
  L.~Fei, S.~Giombi, I.~R.~Klebanov and G.~Tarnopolsky,
  %``Three loop analysis of the critical O(N) models in 6-e dimensions,''
  Phys.\ Rev.\ D {\bf 91}, no. 4, 045011 (2015)
  doi:10.1103/PhysRevD.91.045011
  [arXiv:1411.1099 [hep-th]].
  %%CITATION = doi:10.1103/PhysRevD.91.045011;%%
%\cite{Gracey:2015tta}
\bibitem{Gracey:2015tta}
  J.~A.~Gracey,
  %``Four loop renormalization of $\phi^3$ theory in six dimensions,''
  Phys.\ Rev.\ D {\bf 92}, no. 2, 025012 (2015)
  doi:10.1103/PhysRevD.92.025012
  [arXiv:1506.03357 [hep-th]].

%\cite{Rosten:2008ts, Percacci:2014tfa, Mati:2014xma, Kamikado:2016dvw}
\bibitem{Rosten:2008ts}
  O.~J.~Rosten,
  %``Triviality from the Exact Renormalization Group,''
  JHEP {\bf 0907}, 019 (2009)
  doi:10.1088/1126-6708/2009/07/019
  [arXiv:0808.0082 [hep-th]].
%\cite{Percacci:2014tfa}
\bibitem{Percacci:2014tfa}
  R.~Percacci and G.~P.~Vacca,
  %``Are there scaling solutions in the $O(N)$-models for large $N$ in $d>4$ ?,''
  Phys.\ Rev.\ D {\bf 90}, 107702 (2014)
  doi:10.1103/PhysRevD.90.107702
  [arXiv:1405.6622 [hep-th]].
%\cite{Mati:2014xma}
\bibitem{Mati:2014xma}
  P.~Mati,
  %``Vanishing beta function curves from the functional renormalization group,''
  Phys.\ Rev.\ D {\bf 91}, no. 12, 125038 (2015)
  doi:10.1103/PhysRevD.91.125038
  [arXiv:1501.00211 [hep-th]].
%\cite{Kamikado:2016dvw}
\bibitem{Kamikado:2016dvw}
  K.~Kamikado and T.~Kanazawa,
  %``Nonperturbative RG analysis of five-dimensional O(N) models with cubic interactions,''
  arXiv:1604.04830 [hep-th].
%\cite{Eichhorn:2016hdi}
\bibitem{Eichhorn:2016hdi}
  A.~Eichhorn, L.~Janssen and M.~M.~Scherer,
  %``Critical O(N) models above four dimensions: Small-N solutions and stability,''
  Phys.\ Rev.\ D {\bf 93}, no. 12, 125021 (2016)
  doi:10.1103/PhysRevD.93.125021
  [arXiv:1604.03561 [hep-th]].
%\cite{Hogervorst:2014rta, Hogervorst:2015akt}
\bibitem{Hogervorst:2014rta}
  M.~Hogervorst, S.~Rychkov and B.~C.~van Rees,
  %``Truncated conformal space approach in d dimensions: A cheap alternative to lattice field theory?,''
  Phys.\ Rev.\ D {\bf 91}, 025005 (2015)
  doi:10.1103/PhysRevD.91.025005
  [arXiv:1409.1581 [hep-th]].
%\cite{Hogervorst:2015akt}
\bibitem{Hogervorst:2015akt}
  M.~Hogervorst, S.~Rychkov and B.~C.~van Rees,
  %``Unitarity violation at the Wilson-Fisher fixed point in 4-$\epsilon$ dimensions,''
  Phys.\ Rev.\ D {\bf 93}, no. 12, 125025 (2016)
  doi:10.1103/PhysRevD.93.125025
  [arXiv:1512.00013 [hep-th]].

%\cite{Vasiliev:1981yc, Vasiliev:1981dg, Vasiliev:1982dc}
\bibitem{Vasiliev:1981yc}
  A.~N.~Vasiliev, Y.~Pismak, M. and Y.~R.~Khonkonen,
  %``Simple Method of Calculating the Critical Indices in the 1/$N$ Expansion,''
  Theor.\ Math.\ Phys.\  {\bf 46}, 104 (1981)
  [Teor.\ Mat.\ Fiz.\  {\bf 46}, 157 (1981)].
  doi:10.1007/BF01030844
  %%CITATION = doi:10.1007/BF01030844;%%
%\cite{Vasiliev:1981dg}
\bibitem{Vasiliev:1981dg}
  A.~N.~Vasiliev, Y.~M.~Pismak and Y.~R.~Khonkonen,
  %``1/$N$ Expansion: Calculation of the Exponents $\eta$ and Nu in the Order 1/$N^2$ for Arbitrary Number of Dimensions,''
  Theor.\ Math.\ Phys.\  {\bf 47}, 465 (1981)
  [Teor.\ Mat.\ Fiz.\  {\bf 47}, 291 (1981)].
  doi:10.1007/BF01019296
%\cite{Vasiliev:1982dc}
\bibitem{Vasiliev:1982dc}
  A.~N.~Vasiliev, Y.~M.~Pismak and Y.~R.~Khonkonen,
  %``1/n Expansion: Calculation Of The Exponent Eta In The Order 1/n**3 By The Conformal Bootstrap Method,''
  Theor.\ Math.\ Phys.\  {\bf 50}, 127 (1982)
  [Teor.\ Mat.\ Fiz.\  {\bf 50}, 195 (1982)].
  doi:10.1007/BF01015292
%\cite{Lang:1990ni}
\bibitem{Lang:1990ni}
  K.~Lang and W.~Ruhl,
  %``Field algebra for critical O(N) vector nonlinear sigma models at 2 < d < 4,''
  Z.\ Phys.\ C {\bf 50}, 285 (1991).
  doi:10.1007/BF01474081
  %%CITATION = doi:10.1007/BF01474081;%%
%\cite{Lang:1991kp}
\bibitem{Lang:1991kp}
  K.~Lang and W.~Ruhl,
  %``The Critical O(N) sigma model at dimension 2 < d < 4 and order 1/n**2: Operator product expansions and renormalization,''
  Nucl.\ Phys.\ B {\bf 377}, 371 (1992).
  doi:10.1016/0550-3213(92)90028-A
  %%CITATION = doi:10.1016/0550-3213(92)90028-A;%%
%\cite{Lang:1992zw}
\bibitem{Lang:1992zw}
  K.~Lang and W.~Ruhl,
  %``The Critical O(N) sigma model at dimensions 2 < d < 4: Fusion coefficients and anomalous dimensions,''
  Nucl.\ Phys.\ B {\bf 400}, 597 (1993).
  doi:10.1016/0550-3213(93)90417-N
  %%CITATION = doi:10.1016/0550-3213(93)90417-N;%%
%\cite{Lang:1992pp}
\bibitem{Lang:1992pp}
  K.~Lang and W.~Ruhl,
  %``The Critical O(N) sigma model at dimensions 2 < d < 4: A List of quasiprimary fields,''
  Nucl.\ Phys.\ B {\bf 402}, 573 (1993).
  doi:10.1016/0550-3213(93)90119-A
  %%CITATION = doi:10.1016/0550-3213(93)90119-A;%%
%\cite{Petkou:1994ad}
\bibitem{Petkou:1994ad}
  A.~Petkou,
  %``Conserved currents, consistency relations and operator product expansions in the conformally invariant O(N) vector model,''
  Annals Phys.\  {\bf 249}, 180 (1996)
  doi:10.1006/aphy.1996.0068
  [hep-th/9410093].
  %%CITATION = doi:10.1006/aphy.1996.0068;%%
%\cite{Petkou:1995vu}
\bibitem{Petkou:1995vu}
  A.~C.~Petkou,
  %``C(T) and C(J) up to next-to-leading order in 1/N in the conformally invariant 0(N) vector model for 2 < d < 4,''
  Phys.\ Lett.\ B {\bf 359}, 101 (1995)
  doi:10.1016/0370-2693(95)00936-F
  [hep-th/9506116].
%\cite{Broadhurst:1996ur}
\bibitem{Broadhurst:1996ur}
  D.~J.~Broadhurst, J.~A.~Gracey and D.~Kreimer,
  %``Beyond the triangle and uniqueness relations: Nonzeta counterterms at large $N$ from positive knots,''
  Z.\ Phys.\ C {\bf 75}, 559 (1997)
  doi:10.1007/s002880050500
  [hep-th/9607174].
%\cite{Gracey:2002qa}
\bibitem{Gracey:2002qa}
  J.~A.~Gracey,
  %``Crossover exponent in O(N) phi**4 theory at O(1 / N**2),''
  Phys.\ Rev.\ E {\bf 66}, 027102 (2002)
  doi:10.1103/PhysRevE.66.027102
  [cond-mat/0206098].
%\cite{Wilson:1973jj}
\bibitem{Wilson:1973jj}
  K.~G.~Wilson and J.~B.~Kogut,
  %``The Renormalization group and the epsilon expansion,''
  Phys.\ Rept.\  {\bf 12}, 75 (1974).
  doi:10.1016/0370-1573(74)90023-4
%\cite{Kleinert:1991rg}
\bibitem{Kleinert:1991rg}
  H.~Kleinert, J.~Neu, V.~Schulte-Frohlinde, K.~G.~Chetyrkin and S.~A.~Larin,
  %``Five loop renormalization group functions of O(n) symmetric phi**4 theory and epsilon expansions of critical exponents up to epsilon**5,''
  Phys.\ Lett.\ B {\bf 272}, 39 (1991)
  Erratum: [Phys.\ Lett.\ B {\bf 319}, 545 (1993)]
  doi:10.1016/0370-2693(91)91009-K, 10.1016/0370-2693(93)91768-I
  [hep-th/9503230].
%\cite{Penedones:2015aga}
\bibitem{Penedones:2015aga}
  J.~Penedones, E.~Trevisani and M.~Yamazaki,
  %``Recursion Relations for Conformal Blocks,''
  arXiv:1509.00428 [hep-th].
  %%CITATION = ARXIV:1509.00428;%%
\end{thebibliography}
\end{document}